\documentclass[11pt]{article}

\usepackage{amsmath}
\usepackage{amssymb}
\usepackage{float}
\usepackage[T1]{fontenc}
\usepackage{graphicx}
\usepackage[utf8]{inputenc}
\usepackage{mathtools}
\usepackage{wasysym}
\usepackage[svgnames]{xcolor}
\usepackage[backend=biber, sorting=none]{biblatex}
\usepackage[paperheight=27.94cm,paperwidth=21.59cm,left=2.54cm,right=2.54cm,top=2.54cm,bottom=2.54cm]{geometry}

\title{Interpreting Sequence-Levenshtein distance for determining error type and frequency between two embedded sequences of equal length}

\begin{document}
\maketitle

\vspace{1\baselineskip}
\begin{center}
\textbf{Robert Logan\textsuperscript{1}\textsuperscript{†}\textsuperscript{$\ast$}, Amy Wangsness Wehe\textsuperscript{2}\textsuperscript{†}, Dori C Woods\textsuperscript{3}, Jon Tilly\textsuperscript{3}, Konstantin Khrapko\textsuperscript{3}}
\end{center}

\vspace{1\baselineskip}
\textsuperscript{1}Science and Technology Division, Biology and Bioinformatics Department, Eastern Nazarene College, Quincy, MA 02170

\textsuperscript{2}Health and Natural Sciences Division, Mathematics Department, Fitchburg State University, Fitchburg, MA 01420-2697

\textsuperscript{3}College of Science, Department of Biology, Northeastern University, 330 Huntington Ave, Boston, MA 02115

\vspace{1\baselineskip}
†Co-first authors

$\ast$Corresponding author

\ \ Robert Logan: robert.logan@enc.edu

\vspace{1\baselineskip}
\textbf{Abstract }

Levenshtein distance is a commonly used edit distance metric, typically applied in language processing, and to a lesser extent, in molecular biology analysis. Biological nucleic acid sequences are often embedded in longer sequences and are subject to insertion and deletion errors that introduce frameshift during sequencing. These frameshift errors are due to string context and should not be counted as true biological errors. Sequence-Levenshtein distance is a modification to Levenshtein distance that is permissive of frameshift error without additional penalty. However, in a biological context Levenshtein distance needs to accommodate both frameshift and weighted errors, which Sequence-Levenshtein distance cannot do. Errors are weighted when they are associated with a numerical cost that corresponds to their frequency of appearance. Here, we describe a modification that allows the use of Levenshtein distance and Sequence-Levenshtein distance to appropriately accommodate penalty-free frameshift between embedded sequences and correctly weight specific error types. 

\vspace{1\baselineskip}

\vspace{15\baselineskip}
\section{Introduction}

\ \ \ \ Levenshtein distance (LD) is a widely used edit distance metric  \cite{levenshtein1966binary}. The LD algorithm identifies the number of insertions, deletions and substitutions needed to convert one sequence to another. LD can also assign weights to each error type. A weighted error has a numerical cost that is inversely associated with its appearance frequency. Common applications of LD include natural language processing such as \textcolor[HTML]{222222}{speech recognition, dialect detection, plagiarism exposure and spell checking } \cite{su2008plagiarism}  \cite{wieling2014measuring}  \cite{yulianto2018autocomplete}  \cite{zgank2012predicting}. LD operates under the assumption of fixed sequence length, with the analytical window frame for computing distance including the full length of the two sequences. However, modifications to the LD algorithm allow comparison of embedded sequences, which experience error-induced frameshift, without any additional distance cost as described in the Sequence-Levenshtein distance (SLD) modification\textcolor[HTML]{222222}{ } \cite{buschmann2013levenshtein}. However, SLD cannot accommodate weighted errors. The ability to allow for frameshift without penalty as well as weighted errors has direct relevance in many molecular biology applications. For example, DNA sequencing platforms can introduce a characteristic error profile into the nucleic acid sequence such that certain error types will occur more frequently than others. In these cases, the more frequently occurring error types should have a smaller error weight associated with their cost of distance as to reasonably accommodate them in sequence comparison\textcolor[HTML]{222222}{ } \cite{logan20223gold}. \textcolor[HTML]{222222}{}

\textcolor[HTML]{222222}{\ \ \ \ \ \ \ \ \ \ \ Combining the benefits of weighted errors and frameshift accommodated can be accomplished by interpreting the location and value of the lowest value along the last column and last row in the completed unweighted LD table. In other words, by }interpreting the SLD position and value on the completed LD matrix, the error types and frequency of appearance between the sequences can be determined. This strategy allows for error-specific weights to be added while also accommodating error-induced frameshift without additional penalty. The following mathematical conjecture describes the relationship between error type and frequency and the location of the lowest value along the last column and last row of the LD matrix. Examples are given for each case, illustrating how this information can be used for interpreting the error profile between strings, which can then be used to incorporate weighted LD with frameshift correction allowance into the sequence analysis. 

Let \( (a_{i}, b_{j})\) be the entry in the \( i^{th}\) row and the \( j^{th}\) column of a LD matrix table created by comparing two strings of the same length, \( n\). Then \( (a_{n}, b_{n})\) will be the lowest right-hand entry in the table, positioned at the last row and last column. When an entry in an embedded string of interest is deleted or inserted, there is a frameshift to the left or to the right, while the window of analysis (the frame) remains the same since the lengths of the two strings are uniform. Any elements beyond the string of interest that constitutes its context will then fill in the empty space in the case of a deletion. In the case of an insertion, elements may be pushed outside the analytical window upstream or downstream. The position of each base in the sequence will be denoted as \( [S_{1}, S_{2}, S_{3}\ldots  S_{n-1}, S_{n}]\). A LD matrix table is provided in Figure 1 as a reference for analyzing the changes in the following cases. The green highlighted cells that appear in the case matrices represent the SLD positions, which are the cells to be interpreted. 

\begin{figure}[H]
\centering
\includegraphics[width=10.84cm,height=9.36cm]{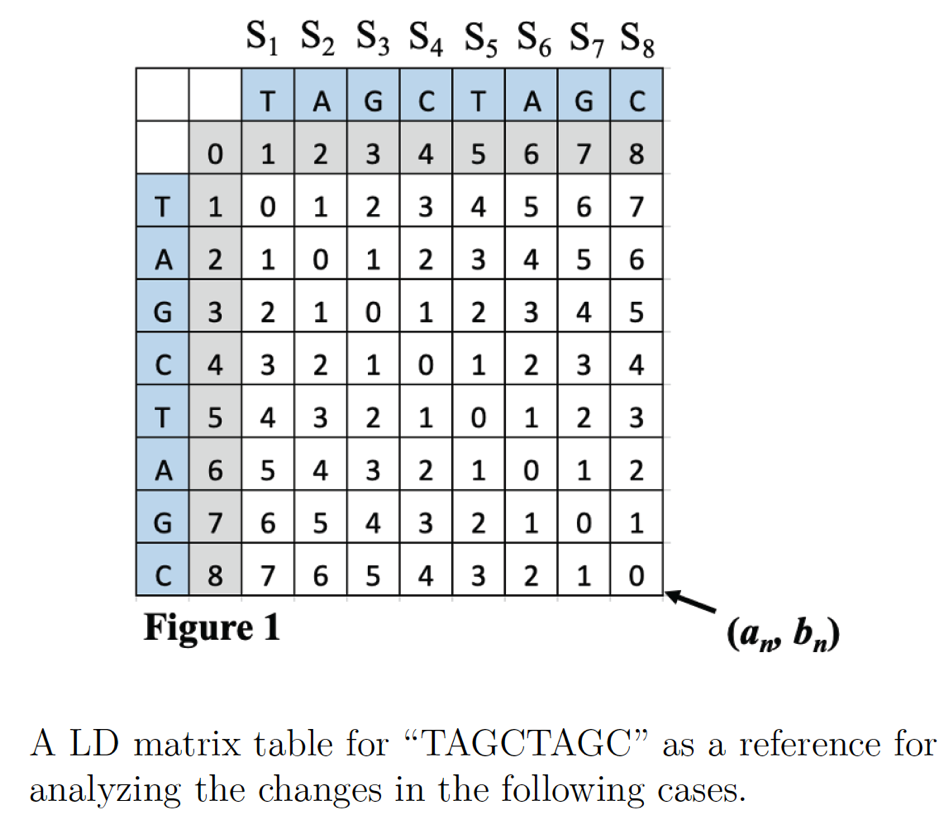}
\end{figure}

\section{Results}

\section{Case 1}

Given \textit{k} insertions, \textit{l} deletions, any number of substitutions, no error at entry \textit{S\textsubscript{n}} regardless of \textit{S\textsubscript{n}} frameshift, and no insertion(s) between \textit{S\textsubscript{n-1 }}and \textit{S\textsubscript{n}\textsubscript{,}} the entry with the lowest value in column \textit{n} and row \textit{n} will be the following:

\begin{equation*}
\left\{ 
\begin{array}{c}
\left(a_{n}, b_{n-\left(l-k\right)}\right)     if l\geq k \\ 
(a_{n-(k-l)},  b_{n})     if k>l \\ 
\end{array}\right.
\end{equation*}

\textbf{\textit{Case 1 example}}

To change ``TAGCTAGC" to ``TAGTAGCT", the operations include a deletion of ``C" and the addition of the ``T" on the 3’ end due to frameshift. The computed LD matrix between these words and the green highlighted SLD placement and value is interpreted to reveal a single deletion as shown below in Figure 2.

\begin{figure}[H]
\centering
\includegraphics[width=12.24cm,height=7.41cm]{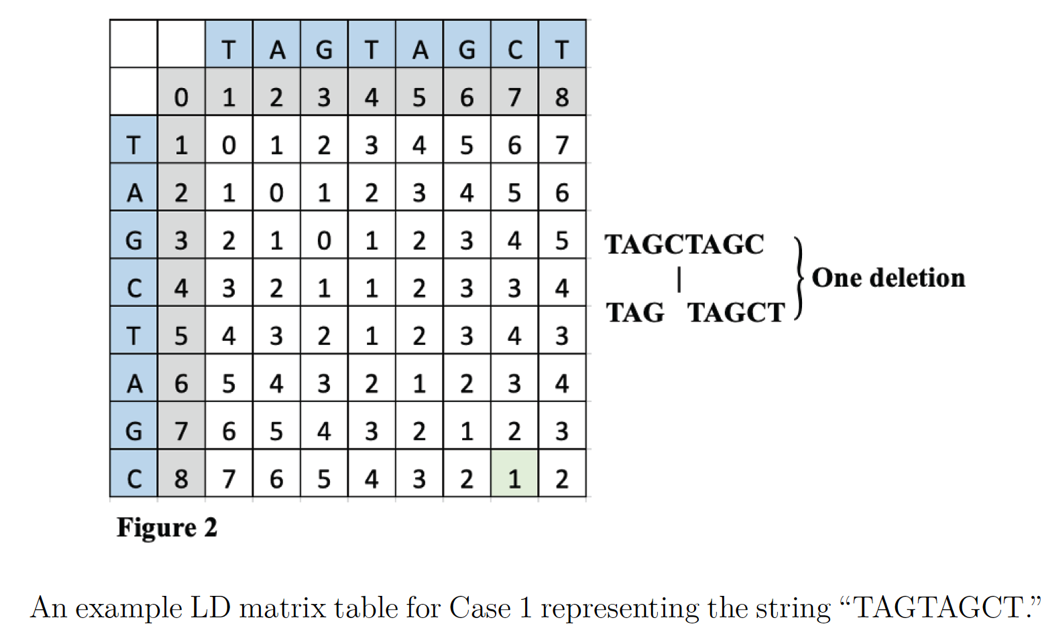}
\end{figure}

\section{Case 2}

Given \textit{d} consecutive deletion-induced, insertion-induced or bona fide substitution(s) that start at \textit{S\textsubscript{n}} and accumulate upstream and no downstream frameshift of the substitutions, regardless of error(s) elsewhere, provided an equal number of insertions or deletions can account for the changes between the sequences, the entries that share the lowest values in column \textit{n} and row \textit{n} will be in the positions \(\left(a_{n}, b_{n-y}\right)\) and \(\left(a_{n-y}, b_{n}\right)\) for all \( y\in\left\{ 0, 1, \ldots , d\right\}\)

\textbf{\textit{Case 2 example}}

To change ``TAGCTAGC" to ``TATAGCTA", the operations can include either a deletion of ``GC" and frameshift towards the 5’ end that allows ``TA" to enter the frame effectively making \textit{S\textsubscript{n}} and \textit{S\textsubscript{n-1}\textsubscript{ }}substituted, or an insertion of ``TA" that pushes ``GC" out of the frame on the 3’ end, also making the \textit{S\textsubscript{n}} and \textit{S\textsubscript{n-1}\textsubscript{ }}substituted. The computed LD matrix between these words and the green highlighted SLD placement and value is interpreted to reveal either two deletions or two insertions as shown in Figure 3. 

\begin{figure}[H]
\centering
\includegraphics[width=13.58cm,height=8.12cm]{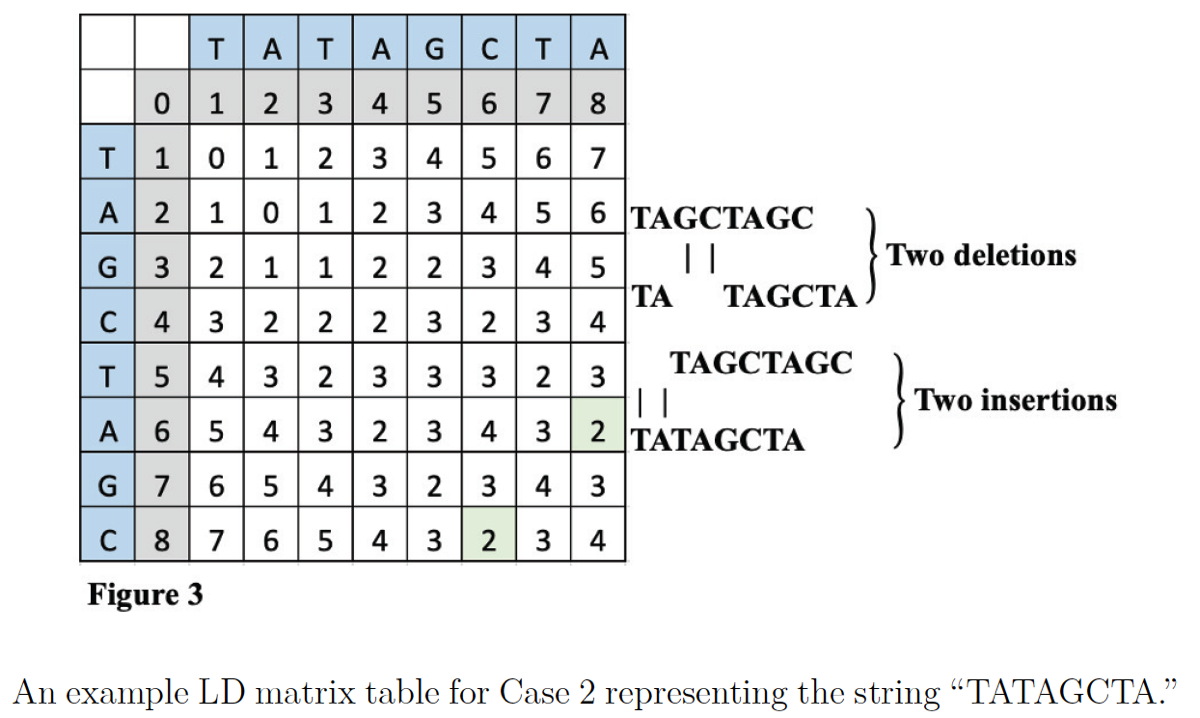}
\end{figure}

\section{Case 3}

Given \textit{d} > 1 consecutive deletion-induced, insertion-induced or bona fide substitution(s) that start at \textit{S\textsubscript{n}} and accumulate upstream and \( p \)insertion-induced downstream frameshifts of the substitutions such that at least one substitution remains at S\textsubscript{n}, regardless of error(s) elsewhere, the entries that share the lowest values in column \textit{n} and row \textit{n} will be in positions \(\left(a_{n-(p)}, b_{n}\right)\) and \(\left(a_{n-(p+k)}, b_{n}\right) \)for \( k\in\left\{ 1,2\ldots y\right\}\). \( y \)\textit{$=$} the number of downstream substitutions left in the analytical window. If frameshift-inducing deletions or insertions neighbor the substitutions, behavior will match case 2.

\section{Case 3 example}

To change ``TAGCTAGC" to ``ATAAGCTG", the operations can include either two insertions of ``A" and one deletion of ``A", two insertions of ``A" and a substitution of ``A" to ``G" or three insertions of ``A". The computed LD matrix between these words and the green highlighted SLD placement and value is interpreted to reveal either two insertions and one deletion, two insertions and one substitution, or three insertions as shown in Figure 4.

\begin{figure}[H]
\centering
\includegraphics[width=14.65cm,height=8.13cm]{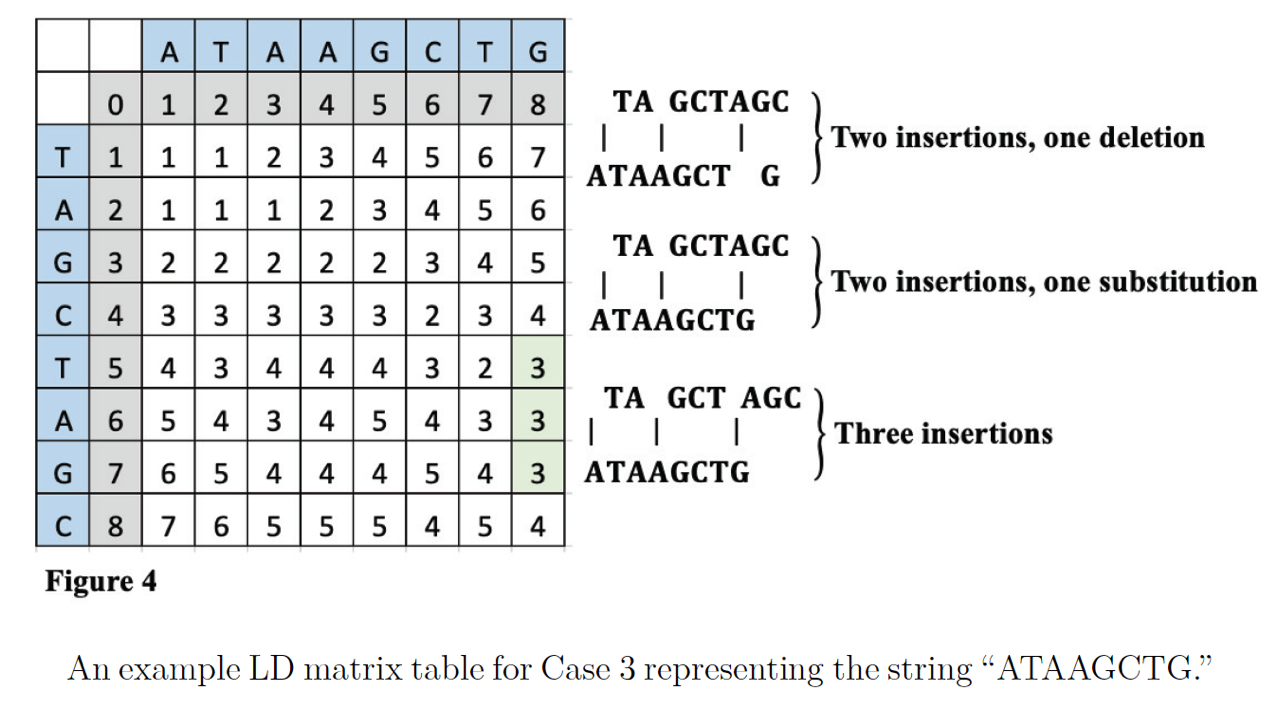}
\end{figure}

\section{Case 4}

Given a \textit{k} $=$ 1 insertion between \textit{S\textsubscript{n-1 }}and \textit{S\textsubscript{n }}and no error at \textit{S\textsubscript{n} or S\textsubscript{n-1}}, any offsetting error(s) elsewhere such that there is no downstream or upstream frameshift of \textit{S\textsubscript{n}}, the entries that share the lowest values in column \textit{n} and row \textit{n} will be \( (a_{n}, b_{n})\), \(\left(a_{n}, b_{n-1}\right)\), and \(\left(a_{n}, b_{n-2}\right)\)

\section{Case 4 example}

To change ``TAGCTAGC" to ``TAGCAGTC", the operations can include either a deletion of ``T" and an insertion of ``T", a deletion of ``T" and substitution of ``C" to ``T", or a deletion of ``T" and ``C". The computed LD matrix between these words and the green highlighted SLD placement and value is interpreted to reveal either an insertion and deletion, a deletion and a substitution, or two deletions as shown in Figure 5.

\begin{figure}[H]
\centering
\includegraphics[width=15.33cm,height=8.1cm]{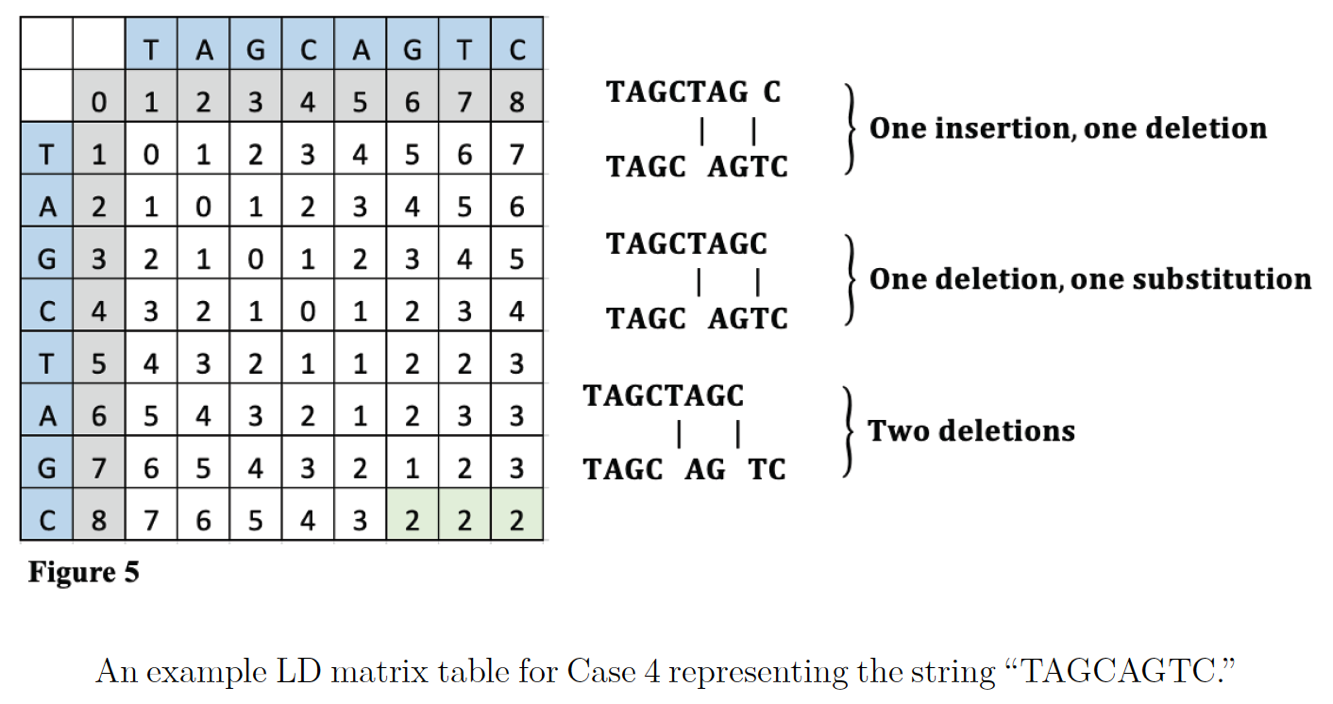}
\end{figure}

\section{Case 5}

Given \textit{l} $=$ 1 deletions of \textit{S\textsubscript{n-1}}, no error at \textit{S\textsubscript{n }}or\textit{ S\textsubscript{n-2}} and no insertion between\textit{ S\textsubscript{n-2}} and \textit{S\textsubscript{n-3}}, and any error(s) elsewhere such that there is no frameshift of \textit{S\textsubscript{n}},\textit{ }the entries that share the lowest values in column \textit{n} and row \textit{n} will be \( (a_{n}, b_{n})\), \( (a_{n-1}, b_{n})\),  and \( (a_{n-2}, b_{n})\)

\textbf{\textit{Case 5 example}}

To change ``TAGCTAGC" to ``TAGTCTAC", the operations can include either an insertion of ``T" and a deletion of ``G" or an insertion of ``T" and a substitution of ``G" to ``C" or an insertion of a ``T" and a ``C". The computed LD matrix between these words and the green highlighted SLD placement and value is interpreted to reveal either an insertion and deletion, an insertion and a substitution, or two insertions as shown in Figure 6.

\begin{figure}[H]
\centering
\includegraphics[width=14.71cm,height=8.21cm]{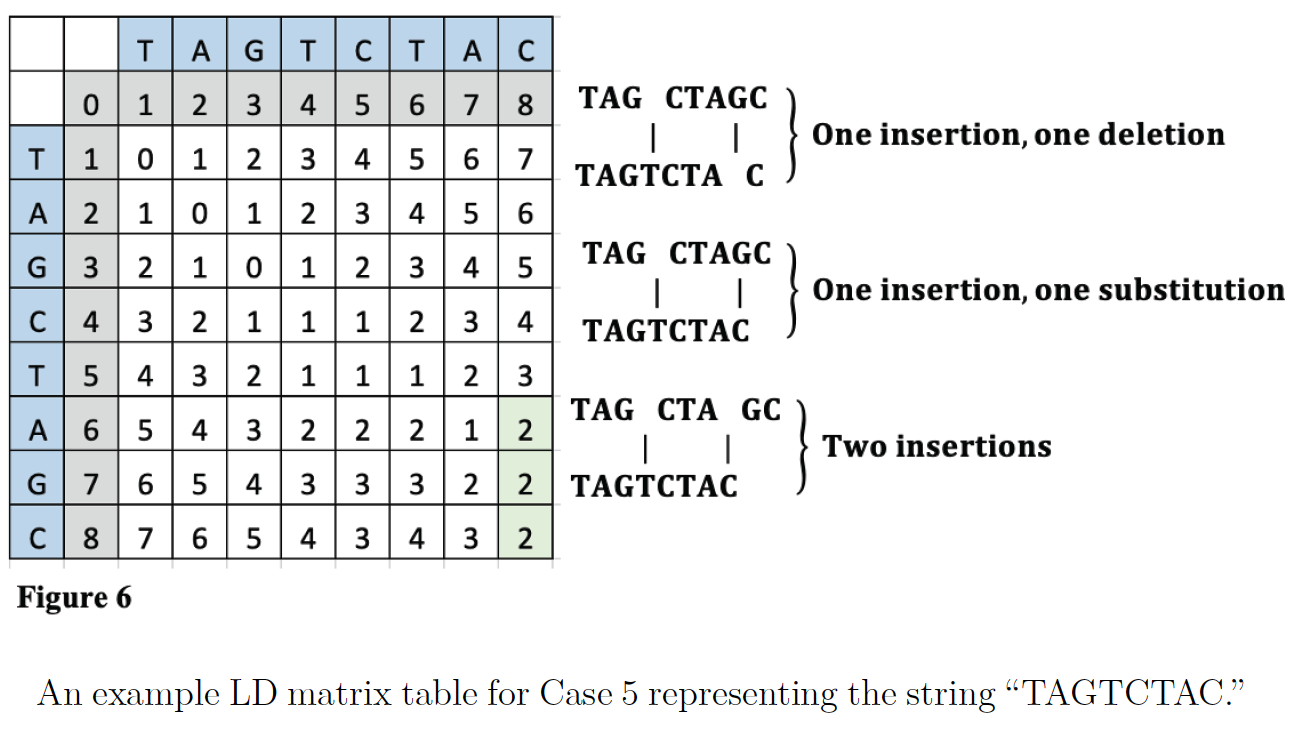}
\end{figure}

\section{Case 6}

Given \textit{l} $=$ 1 deletions of \textit{S\textsubscript{n-1 }}and a substitution error at \textit{S\textsubscript{n} }such that the substitution doesn’t equal the original \textit{S\textsubscript{n-1}}, and any error(s) elsewhere such that there is no frameshift of \textit{S\textsubscript{n}} and the sequences don’t match,\textit{ }the entries that share the lowest values in column \textit{n} and row \textit{n} will be \( (a_{n-1}, b_{n})\),  and \( (a_{n-2}, b_{n})\). If \textit{S\textsubscript{n} }experiences upstream frameshift under these conditions, the lowest values will follow the rules in conjecture 2. If \textit{S\textsubscript{n} }experiences downstream frameshift under these conditions, the lowest values will follow the rules in conjecture 1. 

\section{Case 6 example}

To change ``TAGCTAGC" to ``TAGTCTAT", the operations can include either an insertion of ``T" and a substitution of ``G" to ``T" or an insertion of a ``T" and another ``T". The computed LD matrix between these words and the green highlighted SLD placement and value is interpreted to reveal either an insertion and a substitution or two insertions as shown in Figure 7.

\begin{figure}[H]
\centering
\includegraphics[width=15.21cm,height=8.51cm]{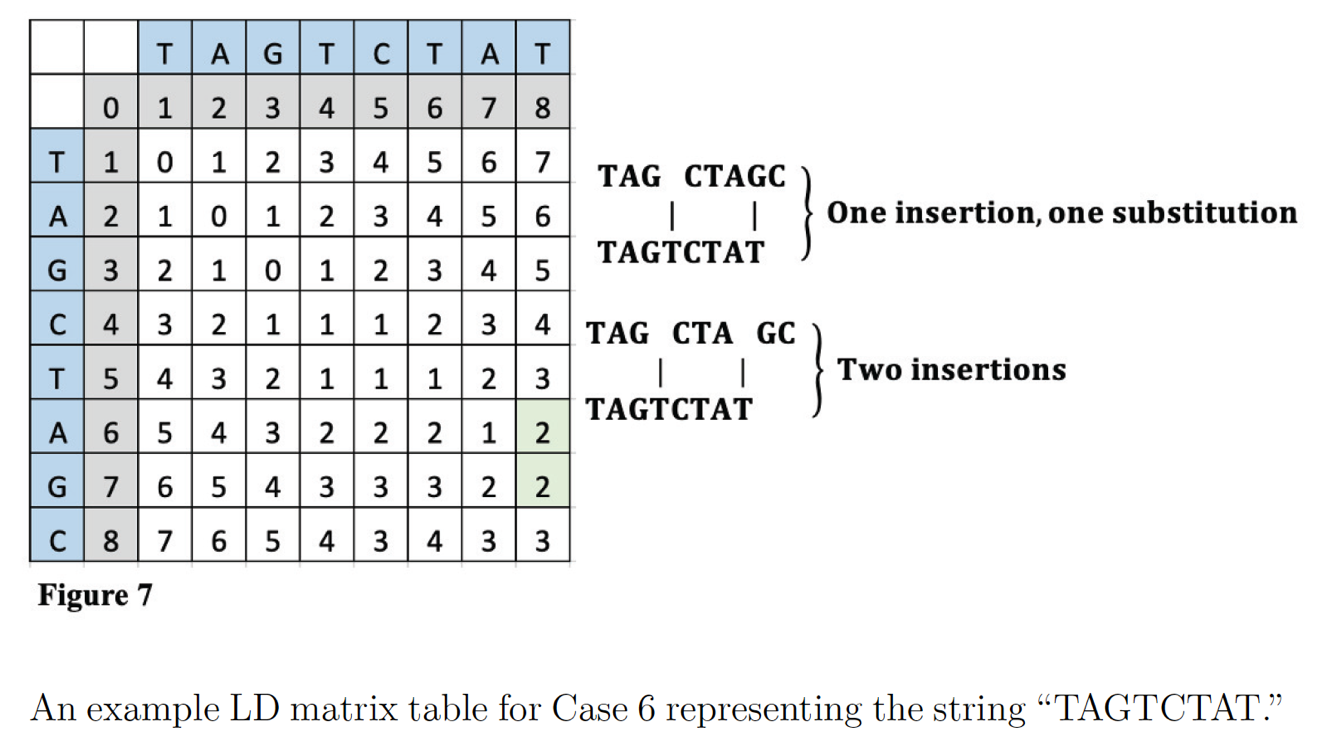}
\end{figure}

\section{Case 7}

Given \textit{l} > 1\textit{ }consecutive deletion errors starting at \textit{S\textsubscript{n-}\textsubscript{1, }}any error(s) elsewhere except consecutive off-setting insertions between \( S_{n-(l+1)}\) and \( S_{n-(l+2)}\) such that there is no frameshift of \textit{S\textsubscript{n}},\textit{ }the entries that share the lowest values in column \textit{n} and row \textit{n} will be \(\left(a_{n-l},b_{n}\right)\) and \(\left(a_{n-\left(l+1\right)},b_{n}\right)\).

\section{Case 7 example}

To change ``TAGCTAGC" to ``ATAAGACC", the minimum number of operations can include insertions of three ``A" letters and a ``C" or insertions of three ``A" letters and a substitution of ``T" to ``C". The computed LD matrix between these words and the green highlighted SLD placement and value is interpreted to reveal either an insertion and a substitution or two insertions as shown in Figure 8. \includegraphics[width=16.51cm,height=8.52cm]{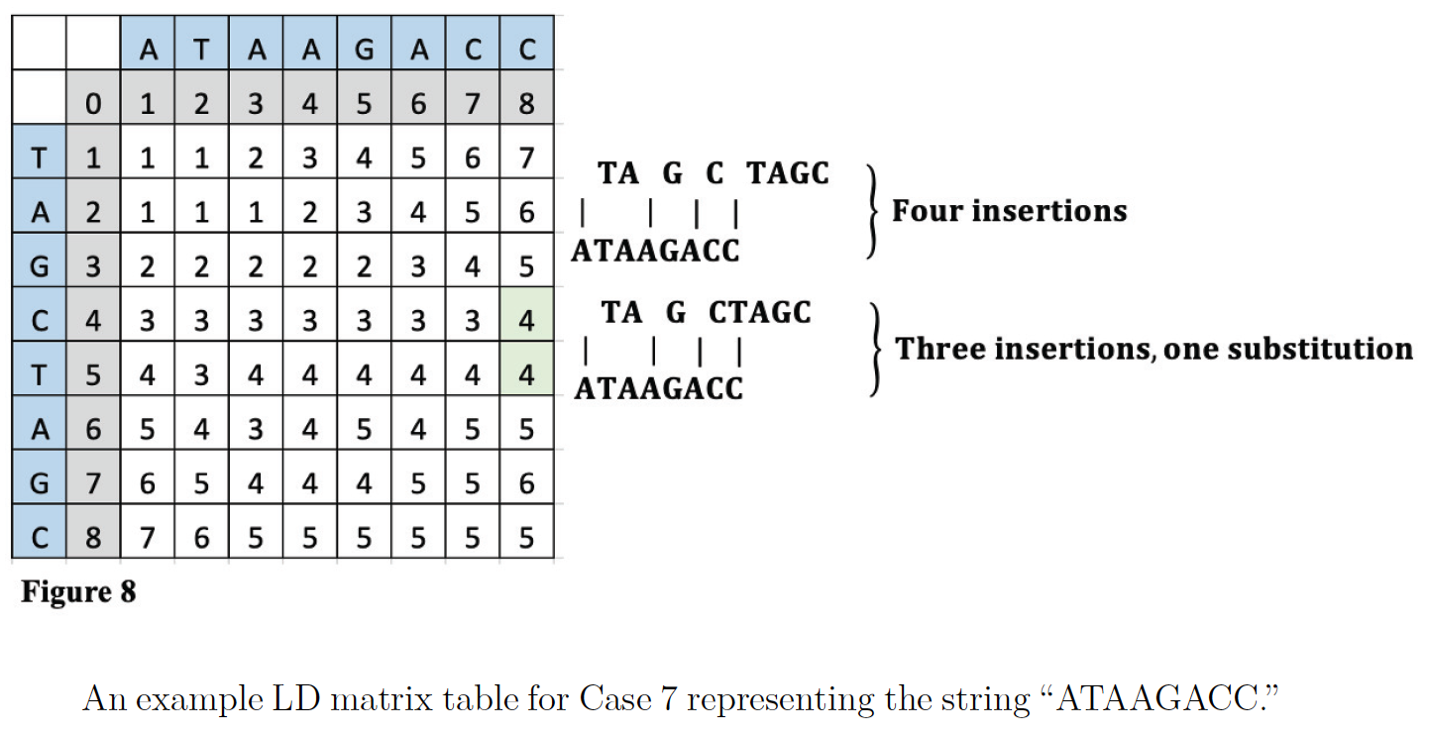}

\vspace{1\baselineskip}
\section{Discussion}

\textbf{\ \ \ \ }The position and value of SLD along the last column and last row of the LD matrix can reveal the error types and appearance frequencies between two sequences of the same length. Insertions move the SLD from the corner up the \( i\) border (the last column) whereas deletions move the SLD from the corner to the left along the \( j\) border (the last row). An insertion matched with a deletion does not move the SLD position along the matrix and neither do substitutions.

Interestingly, an insertion and a deletion pair can produce the same result as a substitution in a sequence if they occur in the same place. However, an insertion and a deletion pair are two errors as opposed to a single substitution error, so distinguishing between these two options matters when counting errors. If an SLD occurs in the corner of the \( i\) and \( j\) borders and the value is 2 or more, it could represent one or more insertion-deletion pairs, all substitutions or a combination of both between sequences. If an SLD occurs in the corner of the \( i\) and \( j\) borders and the value is 1, it represents a single substitution error. Furthermore, some values of two or more in the corner of the \( i\) and \( j\) borders cannot be substitutions, but rather represent only insertion and deletion pairs, as demonstrated in the examples for conjectures 4 and 5. Therefore, if there is ambiguity when interpreting SLD for error type and frequency at the corner of the \( i\) and \( j\) borders, it can be resolved by applying any known probability of whether a substitution is expected to occur more, less or the same as an insertion-deletion pair. Similarly, conjecture 2 describes a scenario where either all insertions or an equal number of deletions can make one sequence match the other. In order to make a determination of which of the two error types is likely responsible for the sequence change for any given analysis, the expected appearance relationship between them for that specific analysis needs to be known. An example of using known probabilities to guide decision making is relying on a specific DNA sequencer’s error hallmark.  

\vspace{1\baselineskip}
\textbf{Acknowledgements }

We wish to thank Boris Yazlovitsky, Greg Shomo, Mariana Levi and the rest of the Research Computing team at Northeastern University for their support. 

\vspace{1\baselineskip}
\textbf{Funding}

This research was supported by the Eunice Kennedy Shriver National Institute of Child Health and Human Development (RO1-HD091439 to JLT, DCW and KK).

\vspace{1\baselineskip}
\textbf{Declaration of interest }

The authors declare that they have no competing interests. 

\printbibliography

\end{document}